\documentclass{elsart}
\journal{/UASLP-IF-04-001}
\begin{document}
\addtolength{\topmargin}{+120pt}

\begin{frontmatter}

\title{ Neutral Fermion Phenomenology With  Majorana Spinors }

\author{M. Kirchbach$^\ast$
}
\ead{mariana@ifisica.uaslp.mx}
\corauth[cor]{Corresponding author}
\address{ Instituto de Fis{\' {\i}}ca, UASLP,
Av. Manuel Nava 6, Zona Universitaria,\\
San Luis Potos{\'{\i}}, SLP 78240, M\'exico}
\author{ C. Compean}
\address{ Instituto de Fis{\' {\i}}ca, UASLP,
Av. Manuel Nava 6, Zona Universitaria,\\
San Luis Potos{\'{\i}}, SLP 78240, M\'exico}
\ead{cliffor@ifisica.uaslp.mx}
\author{L. Noriega
}
\ead{nohemi-27@hotmail.com}
\address{ Facultad de Fis{\' {\i}}ca, UAZ,
Av. Preparatoria 301, Fr. Progreso,\\
Zacatecas, ZAC 98062, M\'exico}

\begin{abstract}
We ask the question whether neutrino physics
with momentum space Majorana spinors, the eigenvectors of 
the particle--antiparticle conjugation operator, 
$C=i\gamma_2\, K$ (with $K$ standing for complex conjugation),
is different but physics with Dirac spinors.
{}First we analyze properties of Majorana spinors
in great detail. We show that four dimensional, $(4d)$,
Majorana spinors are unsuited for the construction of
a local quantum field because $C$ invariance 
does not allow for a covariant propagation in four spinor dimensions,
a conduct due to  $\gamma_2\,  p\!\!\!/=-p\!\!\!/ ^*\, \gamma_2\,  $.
The way out of this dilemma is finding one more discrete
symmetry that respects $C$ invariance and
gives rise to covariant propagators. 
We construct such a symmetry  in
observing that the parity operator, $\gamma_0$, 
``ladders'' between $(4d)$ Majorana rest frame spinors, 
which takes  us to eight dimensional spinor spaces. 
We build up two types of $(8d)$ spaces-- one with a symmetric-
and an other with an anti-symmetric off diagonal metric and
calculate  traces of single beta-- and  neutrinoless 
double beta $(0\nu\beta \beta )$ decays there.
We find physics with $(8d)$ Majorana spinors in the former
space to be equivalent to physics with Dirac spinors
in four dimensions. In the latter space we make the rare observation 
that in effect of cancellations triggered by the anti-symmetric
off diagonal $(8d)$ metric, the neutrino mass 
drops from the single beta decay trace but 
reappears in $0\nu \beta\beta $,
without the neutrino being massless 
in its free equation-- a curious and in principle  
experimentally testable 
signature for a  non-trivial impact of Majorana framework.


\end{abstract}
\end{frontmatter}

\def\beq{\begin{eqnarray}}
\def\eeq{\end{eqnarray}}


\def\s{\mbox{\boldmath$\displaystyle\mathbf{\sigma}$}}
\def\J{\mbox{\boldmath$\displaystyle\mathbf{J}$}}
\def\K{\mbox{\boldmath$\displaystyle\mathbf{K}$}}
\def\P{\mbox{\boldmath$\displaystyle\mathbf{P}$}}
\def\p{\mbox{\boldmath$\displaystyle\mathbf{p}$}}
\def\hp{\mbox{\boldmath$\displaystyle\mathbf{\widehat{\p}}$}}
\def\x{\mbox{\boldmath$\displaystyle\mathbf{x}$}}
\def\0{\mbox{\boldmath$\displaystyle\mathbf{0}$}}
\def\bv{\mbox{\boldmath$\displaystyle\mathbf{\varphi}$}}
\def\hbv{\mbox{\boldmath$\displaystyle\mathbf{\widehat\varphi}$}}

\def\bg{\mbox{\boldmath$\displaystyle\mathbf{\gamma }$}}

\def\bl{\mbox{\boldmath$\displaystyle\mathbf{\lambda}$}}
\def\br{\mbox{\boldmath$\displaystyle\mathbf{\rho}$}}
\def\1{\mbox{\boldmath$\displaystyle\mathbf{1}$}}
\def\bfhh{\mbox{\boldmath$\displaystyle\mathbf{(1/2,0)\oplus(0,1/2)}\,\,$}}

\def\mn{\mbox{\boldmath$\displaystyle\mathbf{\nu}$}}
\def\amn{\mbox{\boldmath$\displaystyle\mathbf{\overline{\nu}}$}}

\def\mne{\mbox{\boldmath$\displaystyle\mathbf{\nu_e}$}}
\def\amne{\mbox{\boldmath$\displaystyle\mathbf{\overline{\nu}_e}$}}
\def\rlh{\mbox{\boldmath$\displaystyle\mathbf{\rightleftharpoons}$}}

\def\wm{\mbox{\boldmath$\displaystyle\mathbf{W^-}$}}
\def\hh{\mbox{\boldmath$\displaystyle\mathbf{(1/2,1/2)}$}}
\def\h00h{\mbox{\boldmath$\displaystyle\mathbf{(1/2,0)\oplus(0,1/2)}$}}
\def\znbb{\mbox{\boldmath$\displaystyle\mathbf{0\nu \beta\beta}$}}



\vspace{1truecm}

\section{Introduction.}  
Virtual exchange of fermions among matter fields is a 
qualitatively new concept in contemporary particle
physics and appears in supersymmetric theories as a process
supplementary to the exchange of ordinary bosonic gauge fields.

Virtual fermions like photino, gravitino etc  
transport interactions and are truly neutral \cite{SUSY}.

{}Further important process of the above type 
is the virtual neutrino line connecting two 
$W^-_\mu e^-$- currents that provides
the major contribution to the spectacular
neutrinoless double beta decay \cite{Kaiser}, \cite{Klapdor}
where lepton number conservation appears violated.

\noindent
Truly neutral fermions, in being their own anti-particles,
are invariant under particle--anti-particle (charge) 
conjugation, $C$, and carry well defined 
$C$ parity, while charged fermions  are 
invariant under space reflection and are endowed with spatial,
$P$, parity. As long as $C$ and $P$ do not commute,
charged and truly neutral fermions are essentially different
species.

Genuinely neutral spin-1/2 fermions are referred to as
Majorana particles, while the charged ones are
the Dirac particles. 

\noindent
The theory of Majorana fermions is based upon quantum fields
that are $C$ eigenstates.
The calculus of widest use for neutral spin 1/2 fermions
is based upon the so called Majorana quantum field,
to be denoted by $\nu (x)$ in the following. 
Its construction is inspired by the Dirac field,
\beq
\Psi_D (x) =\int \frac{d^3\p}{
(2\pi)^{\frac{3}{2}} \sqrt {2p_0}
}
\sum_{h}
{\Big[} 
u_h(\p )  a_{ h }(\p) e^{-i p\cdot x}          
&+&
v_h(\p)\,\,\,
b^\dagger_{ h } (\p)e^{i p\cdot x}
{\Big]}\, .
\label{Dir_field}
\eeq
In defining the transformation properties of
the Dirac spinors (in the convention of Ref.~\cite{KA}) 
and the Fock operators under particle--anti-particle conjugation 
as \cite{Peskin}
\beq
C u_h(\p )=\beta v_{-h}(\p )\, ,&\quad &
Ca_h (\p )C^{-1} = \beta^{-1} b^\dagger_{-h} (\p) \, ,\nonumber\\
 \beta  &=&\delta_{h\downarrow }-\delta_{h \uparrow }\, , 
\label{C_propr}
\eeq
 one concludes for $C$
\beq
C= i\gamma_2 K\, ,
\label{C_C_O}
\eeq
with $K$ standing for the operation of complex conjugation.
The Majorana quantum field is now defined \cite{em1937} in
identifying in Eq.~(\ref{Dir_field}) the operators of 
particle-- and  anti-particle creation  according to
\beq
\nu (x)&=&
\int \frac{d^3\p}{ (2\pi)^{\frac{3}{2}}\sqrt{2 p_0}} 
\sum_{h}
{\Big[} 
u_h(\p ) a_{ h }(\p) e^{-i p\cdot x}          
+ \lambda\, 
v_h(\p)\,\,\,
a^\dagger_{h} (\p)e^{i p\cdot x}
{\Big]}\, ,
\label{field_opr}
\eeq
where $h$ stands for helicity, and
$\lambda $ is the so called {\it creation phase factor\/}
introduced in \cite{KayserPRD}.
The freedom of having the $\lambda $ phase in Eq.~(\ref{field_opr})
is important for obtaining a real mixing matrix under 
$CP$ conservation.
An other option for a neutral quantum field, termed to as $\mu (x)$ by us,
would be to use in place of massive Dirac spinors, which are parity, $P$,
eigenvectors, the massive eigenvectors of the 
particle--anti-particle (charge) conjugation operator, $C$. 
Such spinors are known as momentum space Majorana spinors,  
and are denoted by us as $\Psi_M^{h;(\epsilon_j)}(\p)$. Here,
$\epsilon_j=\pm 1$, or $\pm i$,  fixes $C$ parity.  
A $\mu (x)$ field built upon, say, real $C$ parity momentum spinors
is expected to take the form
\beq
\mu (x)&=&
\int \frac{d^3\p}{(2\pi)^{\frac{3}{2}}\sqrt{2 p_0}} 
\sum_{h}
{\Big[} 
\Psi_M^{h;(+1)} (\p ) a_{ h }(\p) e^{-i p\cdot x}          
+ 
\Psi_M^{h;(-1)}(\p)\,\,\,
a^\dagger_{h} (\p)e^{i p\cdot x}
{\Big]}\, .
\label{field_opr_Maj}
\eeq
Majorana spinors can be encountered
in the neutrino physics chapters of a multitude of
contemporary textbooks such like
\cite{Kaiser}, \cite{Peskin}, \cite{Ramon},
\cite{Ria_Fay}, \cite{Pokorski}.
Now the main question is whether a construct like $\mu (x)$ is reasonable,
and if so, whether it predicts phenomena beyond 
the range of Eq.~(\ref{field_opr}).

\noindent
It is the goal of the present paper
to answer these questions.
We aim to go to the essentials
of the $C$ invariant four-spinors and unveil predictive power
of truly neutral quantum fields entirely based upon
momentum space $C$ parity spinors.
The preprint is organized as follows.
Section 2 reveals various peculiarities of
massive Majorana spinors such as twofold  helicity content
(in the helicity frame), and
self--orthogonality. There we show that $\mu (x)$ is unreasonable
because the  $\Psi_M^{h;( \epsilon_j)}(\p )$'s are non-propagating.
We circumvent the problem of $\p $ independence of the
Majorana propagators in noticing that $p^\mu\gamma_\mu $ ladders between 
certain Majorana spinors, a property that reflects a 
discrete symmetry of Majorana spinors beyond $C$
but in  eight spinorial dimensions, $(8d)$.
We exploit the new discrete symmetries for  the construction
of covariant projectors and corresponding wave equations.

Throughout Section 2 we use the textbook Majorana spinors,
$\Psi_M^{h;(\pm 1 )}(\p )$ of real $C$ parity and
build up the first complete set of eight dimensional spinor 
degrees of freedom.
It is characterized by a metric that is real, off-diagonal,
and symmetric.

In Section 3 we (i) design various $(8d$) currents 
(ii) calculate the neutron beta decay trace,
(iii) find it to be same as if we had worked  in four dimensions 
with Dirac spinors.

We continue in Section 4 with 
Majorana spinors of pure imaginary $C$ parity,
$\Psi_M^{h;(\mp i)}(\p )$ in our notation.
While this type of Majorana spinors has same
gross peculiarities in $(1/2,0)\oplus (0,1/2)$ as the 
textbook ones, it also differs from the latter in
some aspects. In particular, scalar products between such 
Majorana spinors change sign upon reversing 
order of the spinors. This peculiarity shows
up in the associated eight dimensional space as a metric  
that is off--diagonal, purely imaginary and anti--symmetric.

In the latter space we make the rare observation that in effect
of cancellations triggered by the anti-symmetric metric,
the neutral particle mass can drop from the neutron beta decay
trace and one finds a Dirac trace with a {\it massless\/} 
neutral particle sector, an effect that should be in principle 
observable in $\beta $ decay with polarized sources.

In Section 5 we elaborate the trace in the neutrinoless
double beta decay, $0\nu\beta \beta $, by means of
$(8d)$ Majorana spinors and show it to be unaltered with
respect to the standard expression based upon Dirac spinors, 
irrespective of the type of the spinor, or current input.

The paper closes with a brief Summary.

\section{Majorana spinors of real $C$ parity. }

The textbook Majorana spinors (here in momentum space) 
are defined as
\begin{equation}
\Psi^{h;(\epsilon_j)}_{M} (\p )=
\left(
\begin{array}{c}
\epsilon_{j} i\sigma_2 \left(\Phi^{h}_{L}(\p )\right)^*\\
\Phi^{h}_{L}(\p )
\end{array}\right)\, ,
\quad h=\uparrow, \downarrow \, , \quad \epsilon_{1}=-\epsilon_{2}=1\, .
\label{Maj_1}
\end{equation}
Here $\Phi^{h}_{L} (\p )$ is a left handed, $(0,1/2)$, spinor of given
helicity,\footnote{
We prefer to perform our calculations
in the helicity frame where many a statements are more obvious.
Our results are however basis independent, if not emphasized differently.}
$\sigma_2$ is the standard second Pauli matrix, $h=\uparrow, \downarrow $,
and $\epsilon_j$ is the {\it real relative phase \/} 
 between the Weyl spinors that will be identified with
their $C$ parity in the following.

As is well known ~\cite{Peskin}, 
$\s \sigma_{2}=\sigma_{2}(-\s )^*$, and 
\begin{eqnarray}
\hat{\p}\cdot \s \epsilon_j i\sigma_{2}
\, \left( \Phi^{h}_{L}(\p )\right)^*
&=&\epsilon_j\, i\sigma_{2}(- \hat{\p}\cdot \s )^*
\left(\Phi^{h}_{L}(\p )\right)^* \nonumber\\
&=& \epsilon_j\,(- h)\, i\sigma_{2}\left(\Phi^{h}_{L}(\p )\right)^*\, 
=-\epsilon_j \, h\, \alpha\, \Phi^{-h}_R(\p )\, ,
\label{Maj_3}
\end{eqnarray}
meaning that $i\sigma_2 \left(\Phi^{h}_{L}(\p )\right)^*$
is (up to a sign $\alpha  =\pm $),
a right handed field (henceforth denoted by
$\Phi^{-h}_{R}(\p )$) of opposite helicity to $\Phi^{h}_{L}(\p )$.

Therefore, contrary to Dirac spinors,
Majorana spinors can not be prepared as pure helicity eigenstates.
Rather, they are patched together by two Weyl spinors of 
opposite helicities.

\subsection{Rest frame properties.}
The explicit expressions for the rest frame spinors resulting from
Eqs.~(\ref{Maj_1}) and (\ref{Maj_3}) read
\beq
\Psi^{\uparrow ;(+1)}_M(\0 )=
\left(
\begin{array}{c}
\Phi^\downarrow_R (\0 )\\
\Phi_L^\uparrow (\0 )
\end{array}
\right)\, ,&\quad&
\Psi^{\downarrow ;(+1)}_M(\0 )=
\left(
\begin{array}{c}
-\Phi^\uparrow_R (\0 )\\
\Phi_L^\downarrow (\0 )
\end{array}
\right)\, ,\nonumber\\
\Psi^{\uparrow ;(-1)}_M(\0 )=
\left(
\begin{array}{c}
-\Phi^\downarrow_R (\0 )\\
\Phi_L^\uparrow (\0 )
\end{array}
\right)\, ,&\quad&
\Psi^{\downarrow ;(-1)}_M(\0 )=
\left(
\begin{array}{c}
\Phi^\uparrow_R (\0 )\\
\Phi_L^\downarrow (\0 )
\end{array}
\right)\, .
\label{Maj_rf}
\eeq
Preparing rest frame $(1/2,0)$, and $(0,1/2)$ helicity spinors
along the direction of the intended boost is standard  
(compare \cite{Peskin})
\beq
\Phi^\uparrow _{L/R}(\0 )=\sqrt{m} 
\left(
\begin{array}{c}
\cos (\theta/2) e^{-i\frac{\varphi}{2} }\\
\sin (\theta /2 ) e^{i\frac{\varphi}{2} }\,  
\end{array}\right)\, , &&
 \Phi^\downarrow _{L/R}(\0 )=\sqrt{m} \left(
\begin{array}{c}
\sin (\theta/2)e^{-i\frac{\varphi}{2} }\\
-\cos  (\theta /2 ) e^{i\frac{\varphi} {2}} \,  
\end{array}
\right)\, .
\label{rest_fr_sp}
\eeq
We used the convention 
$i\sigma_2\Phi^\uparrow_{L/R}(\0 )=\Phi^\downarrow _{R/L}(\0 )$,
and denoted azimuthal and polar angles  by
$\varphi$, and $\theta $, respectively.

It verifies directly that $\Psi_M^{\uparrow ;(+1)}(\0 )$,
and $\Psi_M^{\downarrow ;(+1)}(\0 )$, are 
of positive--, while
$\Psi_M^{\uparrow ;(-1)}(\0 )$,
and $\Psi_M^{\downarrow ;(-1)}(\0 )$ are of
of negative $C$ parity, 
\beq
C\Psi_M^{h;(+1)}(\0 ) &=&\Psi_M^{h;(+1)}(\0 )\, ,
\nonumber\\
C \Psi_M^{h;(-1)}(\0) &=&-\Psi_M^{h;(-1)}(\0 )\, . 
\label{weird_orthog}
\eeq
In other words, $\epsilon_j=\pm 1 $ can be viewed as a $C$ parity
label.
The essential difference to Dirac spinors is that  
$\Phi^{h}_{R}(\p )\not= i\sigma_2 \left(\Phi^{h}_{L}(\p )\right)^*$.
It is that very difference
which  makes Majorana spinors so special and gives rise to several 
weird  peculiarities, to be explored in the following.

\subsection{Cross-Normalized  Majorana spinors.}

A curiosity occurs when calculating scalar products of the above
Majorana spinors. In first place, all $\Psi_M^{h;(\epsilon_j)}(\0 )$
are self-orthogonal. Second, also spinors of equal $C$
parities happen to be orthogonal. The only non-vanishing scalar 
products are those between  Majorana spinors of opposite
$C$ parities and opposite $h$ labels,
\beq
\overline{\Psi}_M^{\uparrow ;(+1)}(\0 )\Psi_M^{\downarrow ;(-1)}(\0)=
\overline{\Psi}_M^{\downarrow ;(-1)}(\0 )\Psi_M^{\uparrow ;(+1)}(\0 )&=&2m\, ,
\nonumber\\
\overline{\Psi}_M^{\downarrow ;(+1)}(\0 )\Psi_M^{\uparrow ;(-1)}(\0)=
\overline{\Psi}_M^{\uparrow ;(-1)}(\0 )\Psi_M^{\downarrow ;(+1)}(\0 )&=&-2m\
\, ,
\label{bi_orth_real}
\eeq 
with
\beq
\overline{ \Psi}_M^{h; (\epsilon_j)}(\0 )=
\left(
\Psi_M^{ h; (\epsilon_j )} (\0 )\right)^
\dagger \gamma_0\, ,
&\,\,\,& 
\gamma_0 =
\left(
\begin{array}{cc}
0_2 & 1_2\\
1_2 & 0_2
\end{array}
\right)\, . 
\label{lambda_bar}
\eeq
Notice, that  $h$ specifies helicity of the lower
Weyl spinor, $\Phi^h_L(\p )$. 
Majorana spinors were shown above to be of twofold helicity content.
In result, the textbook Majorana spinors build two
independent spaces  distinct by the sign of their 
cross-norms. Each subspace contains a positive-- and a 
negative $C$ parity  spinor of non-vanishing cross--projections
As long as scalar products are Lorentz invariant,
cross-normalization holds true in all inertial frames.

\noindent
The self-orthogonality of Majorana spinors has a devastating
impact on several fundamental operators in
$(1/2,0)\oplus (0,1/2)$ such as mass term and projectors.

\subsection{Absence of a diagonal mass term.}
The structure of a (generic) Majorana spinor, 
$\Psi_M^{h;(\epsilon_j)}$, is more transparent within the 
context of the group $SL(2,C)$ (consult \cite{Hladek} among others), 
where $\Phi_L^h$ is the undotted upper--, while
$i\sigma_2 (\Phi_L^h)^\ast$ is the dotted lower index
spinor.  
We here focus our attention onto the diagonal (Dirac) mass term
$m_M\overline{\Psi}_M^{h;(\epsilon_j)}\Psi_M^{h;(\epsilon_j)}
$. In fact, it can not be introduced at all as 
self-orthogonality nullifies $\overline{\Psi}_M^{h;(\epsilon_j)}
\Psi_M^{h;(\epsilon_j)}$. 

\noindent
One possible way out of vanishing Dirac mass terms 
for Majorana particles proposed in the literature
is to restrict to two component spinors and to consider the Weyl
spinor components,
$\left( \Phi^h_L \right)^*_a $, and 
$\left[ i\sigma_2 \left( \Phi^h_L \right)\right]_b^* $ 
with $a,b=1,2$ as anti--commuting Grassmann numbers.
In so doing, one produces the following mass term \cite{Peskin}
\begin{eqnarray}
m_M
\Phi_L^h\, ^\dagger  
\left[
i\sigma_2 
\left( 
\Phi^h_L
\right)^*
\right]&=&
m_M\left( 
\left(\Phi^h_L  \right)^*_1\, , 
\left(\Phi^h_L \right)^*_2
\right)
\left(
\begin{array}{c}
\left(
\Phi^h_L
\right)^*_2 \\
-\left(
\Phi^h_L 
\right)^*_1
\end{array}
\right)\, ,\nonumber\\
\left(\Phi^h_L \right)^*_1\left(\Phi^h_L \right)_2^*=
&-&
\left(\Phi^h_L \right)^*_2\left(\Phi^h_L \right)_1^*\, .
\label{Grassmann}
\end{eqnarray}
The mass term for $C$ eigenspinors in this scenario
acquires purely quantum nature \cite{Akhmedov}.
It is the first goal of the present study to construct
classical mass terms for $C$ parity spinors.

\subsection{Boosted Majorana spinors.}
In this subsection we consider the effect of 
the $(1/2,0)\oplus (0,1/2)$ boost,
to be referred to as $B_{R\oplus L}(\p )$,
upon $\Psi^{h;(\epsilon_j)}_{M}(\0 )$. 
In making once again use of Eq.~(\ref{Maj_3}) amounts to
\beq
\left(
\begin{array}{c}
\epsilon _{j}\alpha \Phi^{-h}_{R} (\p )\\
\Phi^{h}_{L}(\p )
\end{array}
\right)
&=& \sqrt{\frac{p_0+m}{2\,m}}
\left(
\begin{array}{cc}
1_2 + \frac{\vert \p\vert }{p_0+m}\s\cdot\hat{\p }  & 0_2 \\
0_2 & 1_2 -  \frac{\vert \p\vert }{p_0+m}\s\cdot\hat{\p }
\end{array}
\right)\,
\left(
\begin{array}{c}
\epsilon _{j}\alpha 
\Phi^{-h}_{R} (\0 )\\
\Phi^{h}_{L}(\0 )
\end{array} \right)\, \nonumber\\
&=&\sqrt{\frac{p_0+m}{2\,m}}
\left(1 -h  \frac{\vert \p\vert }{p_0+m}\right)
\left(
\begin{array}{c}
\epsilon _{j}\alpha 
\Phi^{-h}_{R} (\0 )\\
\Phi^{h}_{L}(\0 )
\end{array} \right)\, ,
\label{z}
\eeq
(compare also Ref.~\cite{Ah2}).
Identity and null matrices of dimensionality $n\times n$
are denoted in turn by $1_n$ and $0_n$, while
positive and negative signs in front of the helicity operator,
$\s\cdot\hat{\p} $, correspond to $B_R(\p )$, and $B_L(\p )$, 
respectively,
\cite{Ryder}.
Beyond the representation of the boost matrix 
in Eq.~(\ref{z}), we shall occasionally
use also the following manifestly covariant expressions for
\beq
B_{R\oplus L } (\p )= 
\frac{1}{\sqrt{2m(p_0+m)}} (p\!\!\!/+m\gamma_0)\gamma_0\,  ,
\label{boost_2}
\eeq
and its inversed
\beq
 B_{R\oplus L } (\p )^{-1}= 
\frac{1}{\sqrt{2m(p_0+m)}}\gamma_0\,  (p\!\!\!/+m\gamma_0) .
\label{boost_inverse}
\eeq

\subsection{Spatial parity spinors and Dirac equation.}
In order to obtain  $\Psi_M^{h;(\epsilon_j)}(\p )$ propagation 
one can proceed along the line of 
the general construction of wave equations from the discrete
$P$, and $C$ properties of space time.
Recall, that the group-theoretical construction of the Dirac equation starts
with the rest frame projector onto parity-eigenvectors,
\beq
\Pi^\pm _ {\mathcal R}(\0 ) =\frac{1}{2}\left(1_4\pm \gamma_0 \right)\, ,
\label{rfr_par}
\eeq
and the requirement that the spinors carry good spacial parity
according to
\beq
\Pi^+ _ {\mathcal R}(\0 ) u_h(\0 )=u_h(\0 )\, ,
\quad \Pi^- _ {\mathcal R}(\0 ) v_h(\0 )=v_h(\0 )\, , 
\label{rfr_par_st}
\eeq
with ${\mathcal R}$ labeling space reflection.
Notice that the $\gamma_0$ transformation
of the $SL(2,C)$ spinors amounts to reflections
in three space, i.e. to $\x \to -\x$, $\p \to -\p $.
The boosted form of Eq.~(\ref{rfr_par}) reads
\beq
\Pi^\pm_{\mathcal R}(\p ) =
B_{R\oplus L}(\p ){1\over 2}\left( 1_4 \pm   \gamma_0  \right)\, 
B_{R\oplus L}(\p )^{-1}=\frac{p\!\!\!/\pm m}{2m} \, .
\label{DiR_projs}
\eeq
The Dirac equations for the $u$ and $v$ spinors are then 
\beq
B_{R\oplus L}(\p ){1\over 2}\left( 1_4 +   \gamma_0  \right)\, 
B_{R\oplus L}(\p )^{-1}\, 
u_h(\p )&=&u_h(\p )\,  ,\nonumber\\
B_{R\oplus L}(\p ){1\over 2}\left( 1_4 -   \gamma_0  \right)\, 
B_{R\oplus L}(\p )^{-1}\, 
v_h(\p )&=&v_h(\p )\, .
\label{DiR_eq}
\eeq
The solutions of Eqs.~(\ref{DiR_eq}) can be found, among others,
in Ref.~\cite{KA}. The eigenspinors of the totally symmetric real
matrix $\gamma_0$ are exclusively of real parities and the parity equals 
the relative phase between spinor and co-spinor in 
$(1/2,0)\oplus (0,1/2)$ \cite{Apar}. 
In that regard, the general question on the 
relationship between the relative spinor--co-spinor phase
and parity may  be of interest. In other words, how does an imaginary
spinor--co-spinor relative phase, say,
\beq
U^{\pm i}_h(\p )=
\left(
\begin{array}{c}
\pm i \, \Phi^h_R(\p )\\
\Phi_L^h (\p )
\end{array}
\right)
\label{im_par_sp}
\eeq 
relate to parity?
In order to establish such a link we observe that 
\beq
\gamma_0 \left(
\begin{array}{c}
\pm i \, \Phi^h_R(-\p )\\
\Phi_L^h (-\p )
\end{array}
\right)^* =\mp i \left(
\begin{array}{c}
\pm i \, \Phi^h_R(-\p )\\
\Phi_L^h (-\p )
\end{array}
\right)\, .
\label{anal_cont_par}
\eeq
In a sense, $\gamma_0\, K\, {\mathcal R}$ acts as 
``analytical continuation'' of the $SL(2,C) $ parity operator 
and its eigenstates are of imaginary parity.
Leaving aside the problem of a possible group structure
for which $\gamma_0\, K\, {\mathcal R}$ covers 
space reflections, one nonetheless may try to
subject the projector resulting from  Eq.~(\ref{anal_cont_par}) 
to Lorentzian boost to obtain   
\beq
B_{R\oplus L}(\p ){1\over 2}\left( 1_4 \pm i   \gamma_0  K \right)\, 
B_{R\oplus L}(\p ) ^{-1}\, 
U^{\pm i}_h(\p )=U^{\pm i} _h(\p )\, ,
\eeq
which is equivalent to
\beq
(p\!\!\!/\gamma_0  +m ) 
(p\!\!\!/^* +m\gamma_0) 
U^{\pm i} _h(\p )^\ast  \, &=& \mp i 2m (E+m) U^{\pm i} _h(\p )\, ,
\label{im_p_pr}
\eeq   
an equation that (i) invokes imaginary masses and acausal propagation,
(ii) violates Lorentz invariance.
Therefore, one can not expect any relevance of
such ``imaginary spatial parity'' spinors.
This contrasts the case of the charge conjugation operator which
allows for both real and imaginary parities. Moreover,
for Majorana spinors there is no relationship 
between $C$ parity and causality, a reason for which one
needs to consider both real and imaginary $C$ parities
on equal footing.

\subsection{Non-propagating $C$ parity spinors.\/} 
In taking above  path, we first write down 
the rest frame projectors onto real $C$ parities as  
\beq
{\mathcal P}^{\pm}(\0 )&=&
{1\over 2}\left( 1_4 \pm  i\gamma_2 K \right)\, ,
\quad {\mathcal P}^{\pm}(\0 )\Psi_{M}^{h;(\epsilon_j)}(\0 )=
\Psi_{M}^{h;(\epsilon_j)}(\0 )\, .
\label{proj_op}
\eeq

Boosting ${\mathcal P}^\pm (\0 )$ in  Eq.~(\ref{proj_op}) amounts to
\begin{eqnarray}
{\mathcal P}^\pm (\p ) =
B_{R\oplus L}(\p ){1\over 2}\left( 1_4 \pm  i\gamma_2 K \right)\, 
B_{R\oplus L}(\p ) ^{-1} \, . 
\label{ma_eq_0}
\end{eqnarray}
In substituting Eqs.~(\ref{boost_2}), and (\ref{boost_inverse})
for $B_{R\oplus L}(\p )$, and $B_{R\oplus L}(\p )^{-1}$, 
respectively, amounts to
\beq
{\mathcal P}^\pm (\p ) =
\frac{1}{2}\left( 1_4 \pm \frac{1}{2m(E+m)}(p\!\!\!/\gamma_0 +m) i\gamma_2 
(\gamma_0p\!\!\!/^* +m)\right) 
={\mathcal P}^\pm (\0 )\,,
\label{Schock}
\eeq
where one uses $\gamma_2 \, p\!\!\!/=-p\!\!\!/^*\gamma_2\, $.

In other words, one calculates momentum independence of
the Majorana projector.\\

The consequences would be absence of propagation and imposibility
to construct a local $\mu (x)$. 
This serious drawback of the Majorana spinors requires special
attention, a subject of subsection {\it 2.8\/} below.

\subsection{Static versus non-covariant propagation of $(4d)$
Majorana spinors.} 
A further surprise, perhaps even a pathology, associated with
Majorana spinors is that when exploiting
$\Psi_M^{h;(\epsilon_j)}(\p )$ for the construction of projectors
(here denoted by  $P^\pm (\0 )$) 
onto $C$ parity vectors, one finds them in general
to be different but the analytical projector in Eq.~(\ref{proj_op}).
Consider
\beq
 P ^+(\0 ) = \frac{1}{2m}{\Big(}
\Psi_M^{\uparrow ;(+1)}(\0 )\overline{\Psi}_M^{\downarrow ;(-1)}(\0)
&-&
\Psi_M^{\downarrow ;(+1)}(\0 )
\overline{\Psi}_M^{\uparrow ;(-1)}(\0) 
{\Big)}\, , \nonumber\\
 P^-(\0 ) = -\frac{1}{2m}{\Big(}
\Psi_M^{\uparrow ;(-1)}(\0 )\overline{\Psi}_M^{\downarrow ;(+1)}(\0 )
 &-&
\Psi_M^{\downarrow ;(-1)}(\0 )
\overline{\Psi}_M^{\uparrow ;(+1)}(\0 ){\Big)}\, ,
\nonumber\\
P^+ (\0 ) &+& 
 P^-(\0 )=1_4\, .
\label{C_proj}
\eeq
It directly verifies that $ P^+(\0 )$ and $ P^-(\0 )$ in 
turn project onto spinors of positive and negative $C$ parities according to
\beq
 P^+(\0 )\Psi^{h ;(+1)}_M (\0 ) = 
\Psi^{h ;(+1)}_M(\0 )\, ,
&\quad &
P^-(\0 )\Psi^{h; (-1)}_M(\0 ) = 
\Psi^{h;(-1)}_M(\0 )\, .
\eeq
Naively, one expects $ P^\pm (\0 )$ to coincide with
the analytical rest-frame projector 
${\mathcal P}^\pm = \frac{1}{2}(1_4\pm i\gamma_2 K)$
in Eq.~(\ref{proj_op}). This is by far not so.
The reason is that at rest $K\Psi^{h ;(\epsilon_j)}_M (\0 )$
effectively reduces to  
$
\left( \Psi_M^{h;(\epsilon_j)}(\0 )\right)^* =
\widetilde{{\mathcal A}}\Psi_M^{h;(\epsilon_j)}(\0 )\, ,$ and
\beq
\frac{1}{2}(1_4 \pm i\gamma_2 K )\longrightarrow
\frac{1}{2}( 1_4 \pm i\gamma_2 \widetilde{{ \mathcal A}})\, .
\label{mimick_1}
\eeq
where $\widetilde{{\mathcal A}}$ is particular matrix.
Obviously, $\widetilde{{\mathcal A}}$ depends on the particular choice
for the spinors and can not be frame independent
as long as the operator of complex
conjugation{\it  does not allow for a universal matrix
representation\/}. In case of $\Psi_M^{h;(\pm 1)}(\0 )$,
and in the Cartesian frame, $\widetilde{{\mathcal A}}$ is the unit matrix.
{}For the same reason, in general
\beq
\widetilde{{\mathcal A}}\Psi_M^{h;(\epsilon_j)}(\0 ) &=&
\widetilde{{\mathcal A}} B_{R\oplus L} (\p) ^{-1}
B_{R\oplus L}(\p)\Psi_M^{h;(\epsilon_j)}(\0 )\nonumber\\
&\not=& \Big[B_{R\oplus L} (\p)^{-1}\Big]^\ast
\left(\Psi_M^{h; (\epsilon_j)}(\p)\right)^*\, ,
\label{mimick_2}
\eeq
because as a rule one observes the inequality 
\begin{equation}
\widetilde{{\mathcal A}}  
B_{R\oplus L}(\p ) ^{-1} \widetilde{{\mathcal A}}^{-1}
\not= \lbrack B_{R\oplus L}(\p )^{-1} \rbrack ^\ast\, .
\label{ineq_A_K}
\end{equation}
Next we consider projectors, in turn denoted
by $\Pi^+(\0 )$, and $\Pi^-(\0 )$, onto $\Psi_M^{h;(\pm 1)}(\0 )$
vectors of positive and negative cross-norms:
\beq
 \Pi ^+(\0 ) = \frac{1}{2m}{\Big(}
\Psi_M^{\uparrow ;(+1)}(\0 )\overline{\Psi}_M^{\downarrow ;(-1)}(\0)
&+&
\Psi_M^{\downarrow ;(-1)}(\0 )
\overline{\Psi}_M^{\uparrow ;(+1)}(\0) 
{\Big)}\, , \nonumber\\
 \Pi^-(\0 ) = -\frac{1}{2m}{\Big(}
\Psi_M^{\uparrow ;(-1)}(\0 )\overline{\Psi}_M^{\downarrow ;(+1)}(\0 )
 &+&
\Psi_M^{\downarrow ;(+1)}(\0 )
\overline{\Psi}_M^{\uparrow ;(-1)}(\0 ){\Big)}\, ,
\nonumber\\
\Pi^+ (\0 ) &+& 
 \Pi^-(\0 )=1_4\, .
\label{mass_proj}
\eeq
As long as according to Eq.~(\ref{bi_orth_real}) vectors
of equal cross norms are of opposite $C$ parities, 
the projectors $\Pi^\pm (\0 )$ are in general different from
$P^\pm (\0 ).$\footnote{As we shall see below, in choosing a pure imaginary
$(1/2,0)$--$(0,1/2)$ relative phase,  
one at least achieves equality of the projectors onto vectors of 
positive/negative cross-norms and those onto vectors of 
positive/negative $C$ parities but without resolving the problem
of their non-covariance.}
An immediate and quick test of the latter statement is
performed in the Cartesian frame ($\theta =\varphi =0$ in
Eq.~(\ref{rest_fr_sp}) )
where one calculates $\Pi^\pm (\0 )=\frac{1}{2}(1_4\mp \gamma_5 \gamma_1)$,
while $P^\pm (0 )=\frac{1}{2}(1_4 \pm i\gamma_2\widetilde{{\mathcal A}})$
with $\widetilde{{\mathcal A}} =1_4$.
Apparently, both
$B_{R\oplus L}(\p )\frac{1}{2}(1_4\pm i\gamma_2
\widetilde{{\mathcal A}})B_{R\oplus L}(\p )^{-1}$,
and 
$B_{R\oplus L}(\p )\frac{1}{2}(1_4\mp \gamma_5\gamma_1)
B_{R\oplus L}(\p )^{-1}$
give rise to two essentially different non-covariant equations.

Certainly, one may consider the frame
dependent equations and non-local $\mu (x)$
resulting from boosting $P^\pm (\0 )$, and/or  $\Pi^\pm (\0 )$,
and advocate arbitrary violation of Lorentz symmetry in the Universe,
a path pursued in  Ref.~\cite{Ah2}, and for the spinors
in Section 4 below.
We here take a distinct position and aim to search 
for covariant equations that are 
consistent with the boosted projectors. 
We circumvent the problem of static Majorana propagators 
in $(1/2,0)\oplus (0,1/2)$ to the cost 
of introducing auxiliary extra spinor dimensions.
In the latter space we shall establish consistency between the 
covariant equations and the projectors onto the degrees of freedom 
under consideration and shall construct a local Majorana quantum field.

\subsection{Constructing covariantly propagating Majorana spinors.}
In this subsection we develop an idea how to circumvent
the absence of Majorana spinor propagation in $(1/2,0)\oplus (0,1/2)$ 
observed above.
A simple observation sheds
strong light onto the problem under investigation.
In looking onto Eq.~(\ref{Maj_rf}), it is not difficult to
realize that the parity operator, $\gamma_0$, ``ladders'' between
Majorana spinors of opposite charge conjugation parities and
opposite helicities of the source spinor $\Phi^h_L (\0 )$,
according to
\begin{eqnarray}
\gamma_0 \Psi_M^{\uparrow ;(+1)}(\0 )&=&\Psi_M^{\downarrow ;(-1)}(\0)\, ,
\nonumber\\
\gamma_0 \Psi_M^{\downarrow ;(-1)}(\0 )&=&\Psi_M^{\uparrow ;(+1)}(\0)\, ,
\nonumber\\
\gamma_0 \Psi_M^{\downarrow ;(+1)}(\0 )&=&-\Psi_M^{\uparrow ;(-1)}(\0)\, ,
\nonumber\\
\gamma_0 
\Psi_M^{\uparrow ;(-1)} (\0 )&=&-
\Psi_M^{\downarrow ;(+1)}
(\0)\, .
\label{paso_1}
\end{eqnarray}
This observation takes one directly to a new discrete symmetry
in the larger space of eight spinorial dimensions.
The new symmetry is associated with rest
frame projectors and spinors of the type
\beq
\pi^\pm(\0 )= \frac{1}{2}
\left(\, 1_8 \pm   \left(
\begin{array}{cc}
0_4& \gamma_0\\
 \gamma_0 & 0_4
\end{array}
\right)\, \right)\,, \quad
\pi^+(\0 )
\left(
\begin{array}{c}
\Psi_M^{\downarrow ;(-1)}(\0 )\\
\Psi_M^{\uparrow ;(+1)}(\0 )
\end{array}\right) = 
\left(
\begin{array}{c}
\Psi_M^{\downarrow ;(-1)}(\0 )\\
\Psi_M^{\uparrow ;(+1)}(\0 )
\end{array}\right)\, .\nonumber\\
\label{8_space_proj}
\eeq 
In now defining charge conjugation in the enlarged space
as diag$(-i\gamma_2 K, i\gamma_2 K)$, one immediately realizes
that (i) the blown up spinors carry a well defined $C$ parity,
(ii) the $C$ operator commutes with  $\pi^\pm (\0 )$.
We exploit the new discrete symmetry 
for the construction of covariant projectors 
in subjecting $\pi^\pm (\0 )$ 
to similarity transformations by the boost with the following
result:
\beq
\frac{1}{2m}\left(
\begin{array}{cc}
m1_4 &p\!\!\!/\\
p\!\!\!/& m 1_4 
\end{array}
\right)
\left(
\begin{array}{c}
\Psi_M^{\downarrow ;(-1)}(\p )\\
\Psi_M^{\uparrow ;(+1)}(\p )
\end{array}
\right)
= \left(
\begin{array}{c}
\Psi_M^{\downarrow ;(-1)}(\p )\\
\Psi_M^{\uparrow ;(+1)}(\p )
\end{array}
\right)\, .
\label{8_space_1}
\eeq
Similarly, one finds
\beq
\frac{1}{2m}\left(
\begin{array}{cc}
m 1_4 &-p\!\!\!/\\
-p\!\!\!/ & m1_4
\end{array}
\right)
\left(
\begin{array}{c}
\Psi_M^{\uparrow ; (-1)}(\p )\\
\Psi_M^{\downarrow ;(+1)}(\p )
\end{array}
\right)
=
\left(
\begin{array}{c}
\Psi_M^{\uparrow ;(-1)}(\p )\\
\Psi_M^{\downarrow  ;(+1)}(\p )
\end{array}
\right)\, .
\label{8_space_2}
\eeq
Equations (\ref{8_space_1}) and (\ref{8_space_2})
are equivalently rewritten to
\beq
\left(
\begin{array}{cc}
p\!\!\!/ &0_4\\
0_4& p\!\!\!/ 
\end{array}
\right)
\left(
\begin{array}{c}
\Psi_M^{h ; (\epsilon_j)}(\p )\\
\Psi_M^{-h ;(-\epsilon_j)}(\p )
\end{array}
\right)
= \pm m
\left(
\begin{array}{cc}
0_4 &1_4\\
1_4&0_4
\end{array}
\right)
\left(
\begin{array}{c}
\Psi_M^{h ;(\epsilon_j)}(\p )\\
\Psi_M^{-h  ;(-\epsilon_j)}(\p )
\end{array}
\right)\, .
\label{8_space_mt}
\eeq
The off diagonal form of the $(8d)$ mass matrix
in Eq.~(\ref{8_space_mt}) expresses 
cross-normalization of
$\Psi_M^{h ;(\pm 1)}(\p )$, and its symmetric character
reflects the independence of the cross-norm
on the order of the spinors.
Equations (\ref{8_space_1}) and (\ref{8_space_2}) 
demonstrate how  Majorana spinors propagate in eight
dimensions  and that the propagating
degrees of freedom are well represented by
the following complete set of eight dimensional spinors:
\beq
\Lambda_{1}(\p )=\left(
\begin{array}{c}
\Psi_M^{\uparrow ; (+1)} (\p )\\
\Psi_M^{\downarrow ;(-1)}(\p )
\end{array}
\right)\, , &\quad&
\Lambda_{ 2}(\p )=\left(
\begin{array}{c}
\Psi_M^{\downarrow ; (-1)}(\p )\\
\Psi_M^{\uparrow ;(+1)}(\p )
\end{array}
\right)\, , \nonumber\\
\Lambda _{3}(\p )=\left(
\begin{array}{c}
\Psi_M^{\uparrow ;(+1)} (\p )\\
-\Psi_M^{\downarrow ; (-1)} (\p )
\end{array}
\right)\, , &\quad&
\Lambda_{4} (\p )=\left(
\begin{array}{c}
\Psi_M^{\downarrow ; (-1)} (\p )\\
-\Psi_M^{\uparrow ;(+1)}(\p )
\end{array}
\right)\, ,\nonumber\\
\Lambda_{5}(\p )=\left(
\begin{array}{c}
\Psi_M^{\downarrow ;(+1)} (\p )\\
\Psi_M^{\uparrow ;(-1)}(\p )
\end{array}
\right)\, , &\quad&
\Lambda_{ 6}(\p )=\left(
\begin{array}{c}
\Psi_M^{\downarrow ;(+1)}(\p )\\
\Psi_M^{\uparrow ;(-1)}(\p )
\end{array}
\right)\, , \nonumber\\
\Lambda _{7}(\p )=\left(
\begin{array}{c}
\Psi_M^{\downarrow ;(+1)} (\p )\\
-\Psi_M^{\uparrow ; (-1)} (\p )
\end{array}
\right)\, , &\quad&
\Lambda_{8} (\p )=\left(
\begin{array}{c}
\Psi_M^{\downarrow ;(+1)} (\p )\\
-\Psi_M^{\uparrow ;(-1)}(\p )
\end{array}
\right)\, .
\label{d_space}
\eeq
Notice that above spinors define an orthonormal basis  
as
\beq
\bar \Lambda_{1}(\p )\Lambda_{1}(\p )=
\bar \Lambda_{2}(\p )\Lambda_{2}(\p )&=&
\bar \Lambda_{7}(\p )\Lambda_{7}(\p )=
\bar \Lambda_{8}(\p )\Lambda_{8}(\p )= 4m\, ,
\nonumber\\
\bar \Lambda_{3}(\p )\Lambda_{3}(\p )=
\bar \Lambda_{4}(\p )\Lambda_{4}(\p )&=&
\bar \Lambda_{5}(\p )\Lambda_{5}(\p )=
\bar \Lambda_{6}(\p )\Lambda_{6}(\p )= -4m\, , \nonumber\\
\bar \Lambda_{k}(\p )= \lbrack \Lambda_{k}(\p )\rbrack \, 
^\dagger \, \Gamma_8\,\Gamma^0 , &\quad & k=1,...,8, \quad
\Gamma_8 =\left(
\begin{array}{cc}
0_4&1_4\\
1_4&0_4
\end{array}
\right)\, .
\label{m8_real}
\eeq
Here, $\Gamma_8$ plays the role of metric in the eight dimensional space
of the $\Lambda_k(\p )$  spinors. 
Next we check the energy-momentum dispersion relation 
for the $(8d)$ Majorana spinors.
{}For this purpose we cast, say,  Eq.~(\ref{8_space_1}) into the form
\beq
&&\left(
\begin{array}{cc}
-m1_4 &  p\!\!\!/ \\
 p\!\!\!/ & -m1_4
\end{array}
\right) \,
\left(
\begin{array}{c}
\Psi_{M}^{\downarrow ; (-1)}(\p )\\
\Psi_{M}^{\uparrow ;(+1)}(\p )
\end{array}
\right)= 0\, ,
\label{Dir_8}
\eeq
nullify the determinant of the $8\times 8$ matrix
on the lhs, and  find 
 the time-like relation,
$p^2- m^2=0$. Therefore, Eq.~(\ref{Dir_8}) describes
 neutral particles of real mass 
in terms of spinors that are eigenvectors of the 
particle--anti-particle conjugation operator.

\subsection{Consistency of wave equations  and projectors.}
At that stage it is necessary to test consistency
of Eq.~(\ref{Dir_8}) with the projector onto
the $\Lambda_{k}(\p )$ spinors. In the following 
$\pi^+(\p )$ and $\pi^-(\p )$ in turn denote projectors
onto $\Lambda_{k} (\p )$ spinors of positive, 
and negative norms according to:
\beq
\pi^+(\p )={1\over {4m}}
{\Big(}
\Lambda_{1}(\p )\bar \Lambda_{1}(\p )
+\Lambda_{2}(\p )\bar \Lambda_{2}(\p )
&+&\Lambda_{7}(\p )\bar \Lambda_{7}(\p )
+\Lambda_{8}(\p )\bar \Lambda_{8}(\p ){\Big)},\nonumber\\ 
\label{pr_1}\\
\pi^-(\p )=-{1\over {4m}}
{\Big(}\Lambda_{3}(\p )\bar \Lambda_{3}(\p )
+\Lambda_{4}(\p )\bar \Lambda_{4}(\p )
&+&\Lambda_{5}(\p )\bar \Lambda_{5}(\p )
+\Lambda_{6}(\p )\bar \Lambda_{6}(\p ){\Big)}.\nonumber\\ 
\label{pr_2}
\eeq
In terms of $\pi^\pm (\p )$, the wave equation for the
propagating $\Lambda_k(\p )$ spinors reads
\beq
\pi^\pm (\p )\Lambda_{k}(\p )=\Lambda_{k}(\p )\, ,
\quad k=1,...,8\, ,
\label{pr_wveq}
\eeq
where $\pi^+(\p )$ applies to $\Lambda_{1}(\p )$,
$\Lambda_{2}(\p )$, $\Lambda_{7}(\p )$, $\Lambda_{8}(\p )$,
while $\pi^-(\p )$ applies to the rest.
A direct calculation of, say, $\pi^+ (\p )$ leads to
\beq
\pi^+(\p )&=&
{1\over {4m}}
\left(
\begin{array}{cc}
2m(\Pi^+(\0) +\Pi^-(\0))&\Sigma (\p )\\
\Sigma (\p ) & 2m (\Pi^+(\0 )+\Pi^-(\0))
\end{array}
\right)\nonumber\\
&=&
{1\over {2m}}\left(
\begin{array}{cc}
 m 1_4&{1\over 2}\Sigma (\p )\\
{1\over 2}\Sigma (\p ) & m 1_4
\end{array}
\right)\, ,
\label{pi_plus}
\eeq
where we exploited completeness of the 
$\Psi_M^{ h ;(\mp 1) } (\p ) $  degrees of freedom,
and introduced 
$\Sigma (\p ) =\sum_{h; \epsilon_j}\Psi_M^{h ;(\epsilon_j)} (\p )
\bar \Psi_M^{ h;(\epsilon_j)}(\p )$. 
This quantity can be reduced to a combination of
Dirac $u$ and $v$ spinors upon decomposing
the complete set of $\Psi_M^{h ;(\epsilon_j)}(\p )$ spinors
into the complete set of Dirac's  $\{u_h(\p),v_h(\p)\}$ spinors. 
In so doing one encounters
\beq
\left(
\begin{array}{c}
\Psi_M^{ \uparrow ;(+1)  }(\p)\\
\Psi_M^{\downarrow ; (+1) }(\p)\\
\Psi_M^{ \uparrow ;(-1) }(\p)\\
\Psi_M^{ \downarrow ;(-1)  }(\p)
\end{array}
\right)=
\frac{1}{2}\left(
\begin{array}{cccc}
1_4 &  1_4 & -1_4 &  1_4 \\
-1_4 & 1_4 & - 1_4 & -1_4 \\
1_4 & -1_4 & -1_4 & -1_4 \\
1_4 & 1_4 & 1_4 & -1_4 
\end{array}
\right)\,
\left(
\begin{array}{c}
u_{\uparrow}(\p)\\
u_{\downarrow}(\p)\\
v_{\uparrow}(\p)\\
v_{\downarrow }(\p)
\end{array}
\right)\, .
\label{omega}
\eeq
The latter equation shows that a $\Psi_M^{h ;(\pm 1)}(\p )$ spinor is
a linear combination of Dirac's $u$ and $v$ spinors of opposite
parities, as it should be due to the non-commutativity of
the $C$ and $P$ operators.

\noindent
In making use of the decomposition in Eq.~(\ref{omega}),
one calculates $\Sigma (\p )$ to be the sum
of the projectors onto Dirac $u$ and $v$ spinors according to
\beq
\Sigma (\p )&= &u_\uparrow (\p )\bar u_\uparrow (\p )
+u_\downarrow (\p )\bar u_\downarrow (\p )
+v_\uparrow (\p )\bar v_\uparrow (\p )
+v_\downarrow (\p )\bar v_\downarrow (\p )\nonumber\\
&=& (p_\nu +p_{\bar \nu})\cdot \gamma + m_\nu- m_{\bar \nu}\, .
\label{sigma}
\eeq
Here we denoted four momentum and masses of the neutral particles 
and anti-particles by $p_\nu$, $p_{\bar \nu}$,  
$m_\nu$, and  $m_{\bar \nu }$, respectively.

\noindent
Consistency between Eqs.~(\ref{Dir_8}), and (\ref{pr_1})
is warrant through
the equalities  $p_\nu=p_{\bar \nu}$, and the related
$m_\nu =m_{\bar \nu}$ following from $CPT$ symmetry.
Therefore, one finds  $\Sigma =2 p\!\!\!/$ and
\beq
\pi^+(\p )=
{1\over 2m}\left(
\begin{array}{cc}
 m 1_4& p\!\!\!/\\
 p\!\!\!/ & m 1_4
\end{array}
\right)\, , 
\label{final_check_wveq}
\eeq 
that establishes the consistency under discussion.

\noindent
Canonical quantization {\it \'a la\/} Dirac is now straightforward
in the eight dimensional spinor space, denoted by
${\mathcal S}_8$, when introducing the  {\it local \/}  
$\nu_{\lbrace 8\rbrace }(x)$ field operator as 
\beq
\nu_{\lbrace 8\rbrace } (x)=
\int dV 
{\Big[ }
\sum_{k=1,2, 7,8}
\Lambda_{k} (\p ) a_{k}(\p )\, e^{-i p\cdot x} &+&
\sum_{j=3,4,5,6 }\Lambda_{j}(\p ) a^\dagger _{j} (\p ) 
\, e^{i p\cdot x}
{\Big] }\, .
\eeq
Here, $dV$ is the appropriate phase volume.

\section{ Neutron $\beta $ decay in ${\mathcal S}_8$ 
and textbook Majorana quantum field.}
In this section we exploit the local truly neutral
quantum field $\nu_{\lbrace 8\rbrace }(x)$ in the calculation of
$\beta $ decay traces. In due course we shall  
motivate  $\nu (x)$ in Eq.~(\ref{field_opr}).   
Our first goal on that journey will be to take
a close look on neutron $\beta $ decay in ${\mathcal S}_8$.
If one wishes to consider physical processes that
involve both  Dirac and Majorana fermions, one
needs to worry about  matching  dimensions of
both spinor spaces. The simplest  way to harmonize dimensions
is to amplify the Dirac spinors similarly to 
Eqs.~(\ref{d_space}). 
In order to respect orthogonality
of $P$ eigenspinors, one has to keep helicities
same at top and bottom. The Dirac eight-spinors 
introduced in this manner are 
\beq
U_{(j;h^\prime) } (\p )=\left( 
\begin{array}{c}
u_{h^\prime }(\p )\\
\epsilon_j u_{h^\prime } (\p )
\end{array}
\right)\, , \quad
V_{(j;h^\prime )} (\p )=
\left( 
\begin{array}{c}
v_{h^\prime } (\p )\\
\epsilon_j v_{h^\prime } (\p )
\end{array}
\right)\, ,  \quad h^\prime=\uparrow, \downarrow\, ,
\label{8_Dir}
\eeq
respectively.  
To simplify notations we
will suppress from now onward the momentum, $\p $, as argument
of spinors and operators.
In order to calculate cross sections, i.e.
current-current tensors, $G_{\mu\nu}$,
in ${\mathcal S}_8$, one has next to make a choice for the
eight-currents. 
In analogy to the Dirac vector current,
we here construct  
\beq
J^\mu_{k\, (j;h^\prime )} &=&
\bar \Lambda_{k} \Gamma^\mu U_{(j;h^\prime )}\, ,
\quad \Gamma^\mu =\gamma^\mu \otimes 1_2\, ,\nonumber\\
J^{\mu, 8} _{k\, (j;h^\prime )} &=&
\bar \Lambda_{k} \Gamma_8\Gamma^\mu U_{(j;h^\prime )}\, ,
\quad k=1,2,7,8\, .
\label{8_currents}
\eeq
{}Four momentum and mass of the Dirac particle will be in 
turn denoted as $p_1$, and $m_1$.
As long as we are not gauging the theory, but are
writing down {\it ad hoc\/} currents, one may think
of the $(8d)$ model for neutron beta decay presented here as
a ``toy'' model. Yet, as it will be shown below, it will
allow  for some very instructive insights into neutrino phenomenology. 
Above currents are conserved in the $m\to m_1$ limit
and have the property to 
take  the $U_{(j;h^\prime )}$ spinor of positive norm
to  $\Lambda_k(\p )$ of positive norm too.
The current--current tensor for, say,
$J^\mu_{k\, (j;h^\prime )}$, reads
\beq
G^{\mu \nu} &=&{1\over 2}\,
\sum_{ k (j;h^\prime )}{1\over 4}  
\bar \Lambda_{k}  \Gamma^\mu
U_{(j;h^\prime )}\,\left( \bar \Lambda_{k} \Gamma^\nu
U_{(j;h^\prime)} \right)^\dagger\, .
\label{8_trace_1}
\eeq
In making use of the definition of $\bar \Lambda_{k}$ given
in Eq.~(\ref{m8_real}), i.\ e.\ 
\beq
\left(\bar \Lambda_{k}\right)^\dagger  &=&
\left(
\Lambda_{k}^\dagger \Gamma_8 \Gamma^0\right)^\dagger
=  \Gamma^0\, ^\dagger \Gamma_8^\dagger \Lambda_{k}\, , 
\label{zwischen_stufe}
\eeq
and the relation 
$\Gamma^\nu \, ^\dagger \Gamma^0 \, ^\dagger =\Gamma^0\Gamma^\nu $,
Eq.~(\ref{8_trace_1}) amounts to
\beq
G^{\mu \nu}&=&
{1\over 2} \sum_{ k(j;h^\prime )}
{1\over 4}  
\bar \Lambda_{k}  \Gamma^\mu
U_{(j;h^\prime )}\,
\overline{ U}_{(j;h^\prime)}\Gamma^\nu \Gamma_8^\dagger \Lambda_{k}\,.
\label{8_trace_3}
\eeq
In the following we shall introduce the notation $\Pi^D $ as
\beq
\Pi^D =\frac{1}{2m_1}\left(
U_{(1;\uparrow)}\overline{U}_{(1;\uparrow)}+
U_{(2;\downarrow)}\overline{U}_{(2;\downarrow)}\right)=
\frac{ p\!\!\!/_1+m_11_4}{2m_1}\left(
\begin{array}{cc}
 1& 1\\
1& 1
\end{array}
\right)\, .
\label{proj_Dirac_U}
\eeq
Converting Eq.~(\ref{8_trace_3}) to trace is now standard
and results in
\beq
G^{\mu \nu}&=&{1\over 2}
tr {1\over 4} 
\left( \Gamma_8^\dagger\, \left(  4m  \pi^+ \right)\, 
\Gamma^\mu \, \left( 2m_1 \Pi^D \right) \,
\Gamma^\nu \right)\, \nonumber\\
&=&{1\over 4}tr\left(
\begin{array}{cc}
0_4&1_4\\
1_4& 0_4
\end{array}
\right)
\left(
\begin{array}{cc}
m 1_4& p\!\!\!/\\
p\!\!\!/& m 1_4
\end{array}
\right)
\left(
\begin{array}{cc}
\gamma_\mu (p\!\!\!/_1 +m_1)\gamma_\nu &\gamma_\mu
(p\!\!\!/_1+m_1)\gamma_\nu\\
\gamma_\mu (p\!\!\!/_1+m_1)\gamma_\nu &
\gamma_\mu (p\!\!\!/_1 +m_1)\gamma_\nu
\end{array}
\right) \nonumber\\
&=&{1\over 2} tr (p\!\!\!/+m) \gamma^\mu(p\!\!\!/_1+m_1)\gamma^\nu\, .
\label{8_trace_final}
\eeq
Therefore, the trace entering the single beta decay is same
as if we had used $\nu (x)$. In this way we reached a further goal
of our investigation, namely, understand appearance of momentum
space Dirac spinors in the Majorana quantum field.

\section{Majorana spinors of pure imaginary $C$ parity. }
\subsection{Spinor construct.}
In this Section we present rest-frame  
neutral spinors which differ from  
the textbook ones  \cite{Peskin}--\cite{Pokorski} by the
relative phase between the left-- and right handed 
Weyl components.
In the textbook case, the phase was real, in the currently presented one, 
it will be purely imaginary. The imaginary relative phase shows up 
as imaginary $C$ parity.  
Above difference will be of profound importance for the phenomenological 
consequences of the theory
\footnote{The relative phase $\zeta_j$
between the two-dimensional $(1/2,0)$ and $(0,1/2)$ should not be
 confused with Kayser's creation phase factor, $\lambda $. 
While $\zeta_j$ tells something about
how to  stick together $(1/2,0)$ and $(0,1/2)$ to
a four dimensional spinor of the desired transformation
properties under discrete $C$, $P$ space-time symmetries,
the $\lambda $ selects a particular linear superposition 
between four dimensional neutral particle- and anti--particle states.
In the ultra-relativistic limit, $E/\vert \p \vert \to 1$,
when the particle and anti-particle states become predominantly
left- and right-handed, respectively, the $\lambda $ phase
factor can conditionally be viewed as the relative phase
between $(1/2,0)$ and $(0,1/2)$. However, in this case,
the four spinor brakes down  any way and the 
relative phase $\zeta_j$ becomes irrelevant to leading order.}.

\noindent
McLennan \cite{Mc} and Case \cite{Case},
constructed Majorana spinors of negative imaginary 
$C$ parity, while  $C$ spinors of positive imaginary parity
have been introduced in Refs.~\cite{Ah2} and shown to be 
necessary for securing completeness in $(1/2,0)\oplus(0, 1/2)$.
Majorana spinors of that type will
be denoted in turn by $\Psi_M^{h ;(\zeta_j)}(\p)$
with $\zeta_j$ purely imaginary.
The $\Psi_M^{h;(\mp i)}(\p )$'s correspond to 
$\lambda_{\lbrace h,-h\rbrace }^{S/A}(\p )$ in Refs.~\cite{Ah2},
where $S$ and $A$ stand for self--conjugate (positive imaginary 
$C$ parity), and anti-self--conjugate (negative imaginary $C$ parity), 
respectively.
The rest frame spinors are therefore chosen as
\beq
\Psi_M^{ h ; (\zeta_j)}(\0) &=& 
\left(
\begin{array}{c}
\zeta_j  \left[i\sigma_2\right] \,K\, \Phi^h _L(\0) \\
\Phi^h _L(\0)
\end{array}
\right)\,, \quad \zeta_1=-i, \,\, \zeta_2=i\, .
\label{refra}
\eeq
It directly verifies that the above spinors are indeed $C$ 
eigenvectors, i.e.,
\beq
C\Psi_M^{h; (-i)} (\0 )
= \zeta_1\, \Psi_M^{h; (-i)} (\0 )\, ,
\quad C \Psi_M^{h; (+i)} (\0 )
&=& \zeta_2\, \Psi_M^{h; (+i)}   (\0 )\, .
\label{Psi_c_new}
\eeq
It has been shown in Ref.~\cite{Ah2} that following
cross--normalization relations (termed to as {\it bi-orthogonality\/} there)
hold true
\beq
\overline{\Psi}_M^{h; (\mp i)}(\0) \Psi_M^{h;(\mp i)} (\0) = 0\,,
\,\,\,
\overline{\Psi}_M^{h; (\mp i)}(\0) \Psi_M^{-h;(\mp i)} (\0) 
= \pm  2 i m
(\delta _{h\uparrow}-\delta _{h\downarrow})\, .
\label{bi_orth_im}
\eeq
The imaginary norms have been  provoked by the imaginary relative phase
$\zeta_{j}$. 
As a consequence, the cross norms change sign upon inversing order
of the spinors as visible from Eq.~(\ref{bi_orth_im}).
At the present stage this may look as a disadvantage
but on long term it will be of favor in so far as
it will allow for physics different but the one related
to the Majorana spinors in Eq.~(\ref{Maj_1}) where the relative
phase has been chosen to be real.

\noindent
The completeness relation for these $C$ eigenspinors is
now obtained as
\beq
\Pi^S (\0 ) =
-\frac{1}{2 i m}
{\lbrack }
\Psi_M^{\uparrow ;(-i)}(\0 ) 
\overline{\Psi}_M^{ \downarrow ;(-i)}(\0 )
&-&\Psi_M^{\downarrow ;(-i)}(\p) 
\overline{\Psi}_M^{ \uparrow ;(-i) }(\0 )
{\rbrack} \, ,
\nonumber\\
\Pi^A (\0 ) =+
{1\over {2im}}
{\lbrack }\Psi_M^{\uparrow ;(+i) }(\0 ) 
\overline{\Psi}_M^{ \downarrow ; (+i)}(\0 )
&-&\Psi_M^{ \downarrow ; (+i) } (\0 )
\overline{\Psi}_M^{ \uparrow ; (+i)}(\0 )
{\rbrack}\, ,\nonumber\\
\Pi^S (\0 ) +\Pi^A (\0 ) &=&1_4\, .
\label{lc}
\eeq
where $\Pi^S (\0 ) $ and $\Pi^A(\0 ) $ denote in turn the 
rest frame projection operators onto
the self-- and anti-self conjugate neutral spinors.
The $\Pi^S(\0 )$ and $\Pi^A(\0 )$ are simultaneously 
$C$ parity projectors.
\begin{quote}
The first advantage of imaginary 
$C$ parities is to equalize cross-norms 
to $C$ parity projectors.
Recall, in the case of a real
$\Phi_R^{-h}(\p )$--$\Phi_L^h(\p )$ relative phase
considered in Section 2, projectors onto vectors
with same cross-norm did not coincide with
projectors onto $C$ eigenvectors.
\end{quote}
However, as long as  Eq.~(\ref{Schock}) was deduced without
any reference to the particular form of
the Majorana spinors, also for $\Psi_M^{h;(\mp i)}(\p )$
the problem of static projectors in $(1/2,0)\oplus (0,1/2)$
stays same.

In order to circumvent this shortcoming we 
apply once again the procedure outlined in subsection {\it 2.8\/}
to $\Psi_M^{h;(\mp i)}(\p )$ and find the following 
new system of coupled matrix equations\footnote{Eq.~(\ref{Maj_Dirac}) 
has been written down for the first time 
(up to notational differences) in Ref.~\cite{VVD1995} though without 
addressing the argument about
the constancy of the projector onto four dimensional $C$ eigenspinros and 
without exploring anyone of its phenomenological consequences.}
\beq
\left(
\begin{array}{cc}
-m 1_4  & -ip^\mu \gamma_\mu \\
i p^\mu \gamma_\mu & -m 1_4  
\end{array}
\right)
\left(
\begin{array}{c}
\Psi_M^{ \uparrow ; (-i) }(\p) \\
\Psi_M^{ \downarrow ;(-i) }(\p) \\
\end{array}
\right)&=&0\, , \nonumber\\
\left(
\begin{array}{cc}
-m 1_4 &  ip^\mu  \gamma_\mu  \\
-ip^\mu  \gamma_\mu & -m1_4    
\end{array}
\right)
\left(
\begin{array}{c}
\Psi_M^{ \uparrow ; (+i) }(\p) \\
\Psi_M^{ \downarrow ; (+i)}(\p)
\end{array}
\right)&=&0\, .
\label{Maj_Dirac}
\eeq

\subsection{Covariantly propagating Majorana spinors in doubled $\h00h$.}
Equations~(\ref{Maj_Dirac}) suggest once again to amplify
dimensionality of  $C$ eigenspinors 
from four to eight in introducing
\beq
\Lambda_{(S/A ;1)}(\p )=\left(
\begin{array}{c}
\Psi_M ^{\uparrow ; (\mp i ) } (\p )\\
\epsilon_1\Psi_M^{ \downarrow ; (\mp i) }(\p )
\end{array}
\right)\, , &\quad&
\Lambda_{(S/A ; 2)} (\p )=\left(
\begin{array}{c}
\Psi_M^{ \downarrow ; (\mp i ) }(\p )\\
\epsilon_1\Psi_M^{ \uparrow ; (\mp i) } (\p )
\end{array}
\right)\, , \nonumber\\
\Lambda _{(S/A ;3)}(\p )=\left(
\begin{array}{c}
\Psi_M^{ \uparrow ; (\mp i) }(\p )\\
\epsilon_2 \Psi_M ^{\downarrow ; (\mp i) }(\p )
\end{array}
\right)\, , &\quad&
\Lambda_{(S/A ;4)} (\p )=\left(
\begin{array}{c}
\Psi_M^{ \downarrow ; (\mp i ) }(\p )\\
\epsilon_2\Psi_M^{\uparrow ; (\mp i) } (\p )
\end{array}
\right)\, ,\nonumber\\
\epsilon_1&=&-\epsilon_2=1 \, .
\label{d_space_2}
\eeq
Here, the index $S$, or $A$ of the $\Lambda $ spinors
is associated in turn  with the negative, or positive sign
of the imaginary $C$ parity.
The above spinors define an orthogonal basis as
\beq
\bar \Lambda_{(S;1)}(\p )\Lambda_{(S;1)}(\p )&=&
\bar \Lambda_{(S;4)}(\p )\Lambda_{(S;4)}(\p )=\nonumber\\
\bar \Lambda_{(A;2)}(\p )\Lambda_{(A;2)}(\p )&=&
\bar \Lambda_{(A;3)}(\p )\Lambda_{(A;3)}(\p )= + 4m\, , \\
\bar \Lambda_{(S;2)}(\p )\Lambda_{(S;2)}(\p )&=&
\bar \Lambda_{(S;3)}(\p )\Lambda_{(S;3)}(\p )=\nonumber\\
\bar \Lambda_{(A;1)}(\p )\Lambda_{(A;1)}(\p )&=&
\bar \Lambda_{(A;4)}(\p )\Lambda_{(A;4)}(\p )= -4m\, ,\\
\bar \Lambda_{(\tau ;k)}(\p )= 
\lbrack \Lambda_{(\tau ;k)}(\p )\rbrack \, ^\dagger \,
\widetilde{\Gamma}_8\,\Gamma^0 , && \widetilde{\Gamma}_8=
\left(
\begin{array}{cc}
0_4&-i 1_4\\
i1_4 &0_4
\end{array}
\right)\, .
\label{m8_2}
\eeq
Here, $\tau =S, A$, the new matrix $\widetilde{\Gamma}_8$ plays once again 
the role of metric in the eight dimensional space
(to be denoted by ${\widetilde{ \mathcal S}}_8$) but this time
in the space built on top of $\Psi_M^{h ;(\mp i)}(\p )$.

{}From Eq.~(\ref{Maj_Dirac}) one  directly reads off
that the eight-dimensional spinors satisfy
the Dirac like equation
\beq
&&
\left(
\begin{array}{cc}
p^\mu \gamma_\mu & \pm i m 1_4 \\
\mp i m 1_4& p^\mu \gamma_\mu 
\end{array}
\right) \,
\left(
\begin{array}{c}
\Psi_M^{ h ;(\mp i)} (\p )\\
\epsilon_j\Psi_M^{-h ; (\mp i) }(\p )
\end{array}
\right)=0 ,
\label{Dir_8_2}
\eeq
where ``$-m$'' and ``$+m$'' in turn correspond to 
$\Lambda_{(S/A ;k)}(\p )$  of positive and negative norms. 
In nullifying the determinant of the matrix acting upon
$\Lambda_{(S/A ;k)}(\p )$ in Eq.~(\ref{Dir_8_2}), one
finds the standard energy-momentum dispersion relation,
$p^2-m^2=0$. Therefore, Eq.~(\ref{Dir_8_2}) describes
massive neutral particles in terms of spinors
that are eigenvectors of the particle--anti-particle
conjugation operator.

Comparison between Eqs.~(\ref{m8_2}) and (\ref{m8_real}) shows that
the $(8d)$ metric takes different form in depending on the
$C$ parity.
In case the $C$ parity is real, the metric, $\Gamma_8$, 
is real and symmetric,
while in case the above parity is pure imaginary, the metric,
$\widetilde{\Gamma}_8$, is imaginary and anti-symmetric.

The difference between $\Gamma_8$, and  $\widetilde{\Gamma}_8$ 
comes about because for
$\Psi_M^{h ;(\mp  i)}(\p )$ the cross-norms depend on
the order of the vectors as visible from
Eq.~(\ref{bi_orth_im}), while for
$\Psi_M^{h;(\pm 1)}(\p )$ they did not, in accordance to 
Eq.~(\ref{bi_orth_real}).

Above difference is of pivotal
importance for Eq.~(\ref{8_trace_final}).
Had we used $\Lambda_{(\tau ,k)}(\p )$ in place of
$\Lambda_k(\p )$, i.e.
Eq.~(\ref{Maj_Dirac}) in place of Eq.~(\ref{final_check_wveq}),
and substituted into Eq.~(\ref{8_trace_final})
$\widetilde{\Gamma}_8$ from Eq.~(\ref{m8_2}), 
we would have observed a cancellation
of mass in the neutral fermion sector of the trace.
In effect, the neutral particle sector of the (single) beta decay
trace would be massless without the neutrino being 
massless in reality.

A different situation is obtained in considering the current
(it is conserved in the $m\to m_1 $ limit)
\beq
J^{\mu ,\pm} _{(\tau;k)\, ,(j;h^\prime )} =
\bar \Lambda_{(\tau;k)} {1\over \sqrt{2}} 
\left(1_8\pm \widetilde{\Gamma}_8 \right)
\Gamma^\mu U_{(j;h^\prime )}\, .
\label{8_curr_g8}
\eeq
Here, the interference term
\beq
\pm {1\over 2}{\Big(}\bar \Lambda_{(\tau;k)} \Gamma^\mu U_{(j;h^\prime )}\,
\left(\bar \Lambda_{(\tau;k)} 
\widetilde{\Gamma}_8\Gamma^\nu U_{(j;h^\prime )}\right)^\dagger
&+&\bar \Lambda_{(\tau;k)} \widetilde{\Gamma}_8\Gamma^\mu U_{(j;h^\prime )}\,
\left( \bar \Lambda_{(\tau;k)}\Gamma^\nu U_{(j;h^\prime )}\right)^\dagger\,
{\Big)} ,
\nonumber\\
\label{intf_term}
\eeq
contributes  
$
\pm m\gamma^\mu\left( p\!\!\!/_1 +m_1\right)\gamma^\nu\, ,
$ to the trace in  Eq.~(\ref{8_trace_final}).
This happens because 
$\left( 
\bar \Lambda_{(\tau ;k)}\Gamma_8\Gamma^\nu U\right)^\dagger $=
$\overline{U}^\dagger\Gamma^\nu\Lambda_{(\tau ;k)} $ upon accounting for 
$\widetilde{\Gamma}_8^2=1_8$.Therefore, the antisymmetric off diagonal metric
in $\Lambda_{(\tau ,k)}(\p )$ goes completely away from
the matrix providing the trace, and 
phenomenologies with $\Psi_M^{k;(\pm i)}(\p )$
and $\Psi_M^{k;(\pm i)}(\p )$ amount be same again.

\section{The neutrinoless double beta decay $\0\nu\beta\beta$.}
The neutrinoless double beta decay ($0\nu \beta \beta $) 
is a process where two neutrons in a nucleus, $A(Z,N)$, are 
converted into two protons  by the emission of two virtual
$W^-$ bosons
\beq
A (Z,N) \to  A (Z+2,N-2)+ W^- + W^-\, ,
\label{0bb}
\eeq
in such a way  that the two subsequently
emerging $W^-_\mu e^- $ boson-fermion currents,
 appear connected by a virtual neutrino line
(see Ref.~\cite{Kaiser} for details).
This process is associated with a second order
element of the $S$ matrix and the related amplitude,
here denoted by, $T_{0\nu\beta\beta }$, is
given by
\beq
T_{0\nu\beta\beta }=
W^\mu \, W^\eta
\lbrack \bar u_e \gamma_\mu (1+\gamma_5)u_{\nu_e}\rbrack
\lbrack \bar u_e \gamma_\eta (1+\gamma_5)u_{\nu_e}\rbrack\, .
\label{01_bb}
\eeq
In order to bring in the virtual neutrino line one makes use
of the following identity
\beq
\bar u_e \gamma_\eta (1+\gamma_5)u_{\nu_e}&=&
\overline{ \left( (u_e)^c\right)^c}\gamma_\eta (1+\gamma_5)
\left((u_{\nu_e})^c\right)^c\nonumber\\
&=&\bar u_{\nu_e}\, [-\gamma_\mu (1-\gamma_5)]\, v_e\, .
\label{02_bb}
\eeq
The latter expression is obtained in making use
of the relations,
$\gamma_0\gamma_\mu^\ast=\gamma_\mu\gamma_0$,
$\gamma_2\gamma_\mu=-\gamma_\mu^\ast\gamma_2$,
$\gamma_\mu^t=-\gamma_\mu$, the anticommutation relations
between the Dirac matrices, and $t$ labeling the transposed.
With that Eq.~(\ref{01_bb}) takes the form
\beq
T_{0\nu\beta\beta }&=&
W^\mu \, W^\eta
\frac{1}{p_{\nu_e}^2-m_{\nu_e}^2} 
L_{\mu\eta}\, ,\nonumber\\
L_{\mu\eta}&=&
\, \bar u_e \gamma_\mu (1+\gamma_5)
\Pi^{\nu_e} \, [-\gamma_\mu (1-\gamma_5)]\, v_e \, ,\quad
\Pi^{\nu_e}=\sum u_{\nu_e}\bar u_{\nu_e}\, .
\label{03_bb}
\eeq
Here we suppressed helicity labeling of the Dirac spinors
in order not to overload notations but 
$\sum $ in $\Pi_{\nu_e}$ expresses summation over this
degree of freedom.
{}Finally, $|L_{\mu\eta}|^2$ can be converted to a
trace in the standard way as
\beq
|L_{\mu\eta }|^2 &=&
{\Big[}\bar u_e \gamma_\mu (1+\gamma_5)\Pi^{\nu_e}
\, \gamma_\eta (1-\gamma_5)\, v_e {\Big]}\, 
{\Big[}
\bar u_e \gamma_\lambda (1+\gamma_5)
\Pi^{\nu_e}
\, \gamma_\delta (1-\gamma_5)\, v_e {\Big]}^\dagger\, 
\nonumber\\
&=&tr{\Big[}\Pi^{u_e}\gamma_\mu (1+\gamma_5)\Pi^{\nu_e}\gamma_\eta(1-\gamma_5)
\Pi^{v_e}(1+\gamma_5)\gamma_\delta
\gamma_0 \Pi^{\nu_e}\gamma_0(1-\gamma_5)\gamma_\lambda
{\Big]}\, .\nonumber\\
\label{04_bb}
\eeq
Now we calculate above trace within the scenario
of the previous section. To do so one has to
perform in Eq.~(\ref{04_bb}) the replacements $\gamma_\mu\to \Gamma_\mu$,
$u_e\to U_e$, $v_e\to V_e$, $u_{\nu_e}\to \Lambda_{(S/A; k)}$, and
\beq
\Pi^{\nu_e}\to
\frac{1}{2m}\left(
\begin{array}{cc}
m1_4&-ip\!\!\!/\\
ip\!\!\!/&m1_4
\end{array}
\right)
\left(
\begin{array}{cc}
0_4&-i 1_4\\
i1_4&0_4
\end{array}
\right)\, .
\label{05_bb}
\eeq
In this way one creates the $8\times 8$ version of 
$|L_{\mu\eta }|^2$, where apparently,  
the metric matrix $\widetilde{\Gamma}_8$ enters twice.
The net effect of the $\widetilde{\Gamma}_8^2$ presence
in (\ref{04_bb}) is to bring back the mass to
the neutral particle sector in the $0\nu \beta \beta $ trace. 
Recall, that 
for the  type of currents in Eq.~(\ref{8_currents}) 
and same $\widetilde{\Gamma}_8$,
the neutral particle sector in the 
single $\beta $ decay trace was massless.
Above considerations allow to conclude that
$0\nu \beta \beta $ phenomenology with Majorana spinors 
results equivalent to phenomenology with Dirac spinors.

\section{Summary.}
We demonstrated momentum independence of
the projectors onto  classical $C$
eigenspinors in $(1/2,0)\oplus (0,1/2)$ and concluded
impossibility of constructing local quantum field theory based upon
four dimensional Majorana spinors. 
We avoided the problem of static propagators in $(1/2,0)\oplus (0,1/2)$
in exploiting the fact that in auxiliary eight spinorial dimensions 
Majorana spinors possess  one more discrete symmetry 
beyond charge conjugation. 
This directed us to the auxiliary calculus 
for classical Majorana spinors of eight spinorial degrees of
freedom. In reference to the new symmetry we constructed
related rest-frame projectors which upon boosting gave rise to
covariant propagation and allowed for
a field quantization {\it \'a la Dirac\/}.

\noindent
With the aim to figure out similarities and differences between
Majorana and Dirac theories for neutral fermions,
we calculated the $(8d)$ trace entering the width
of neutron $\beta $ decay within such a scenario
and, up to one exception, found it
to be same as if we had used  in four space massive 
Dirac spinors.

We also calculated the trace entering the neutrinoless double
$\beta $ decay and found it to be unaltered with respect to
the Dirac formalism irrespective of the type of spinors
and type of currents used. 
In other words, we showed that phenomenology
with classical Majorana spinors is possible only in eight spinorial
dimensions, but is by and large equivalent to phenomenology with 
Dirac's $u$ and $v$  in four spinorial dimensions.

If this were to be the only impact of the calculus,
eight spinor dimensions could be viewed only as dummy
degrees of freedom.
However, there is a rare exception. 
{}For $\Psi_M^{k;(\pm i)}(\p )$ 
and the class of currents in Eq.(\ref{8_currents})
the single beta decay trace was shown to contain
massless Dirac spinors in the neutral fermion sector.
This cancellation of the neutral particle
mass was triggered by the anti-symmetric off diagonal metric
in the  $(8d)$ space.  
The latter option opens the curious possibility to have a 
neutral fermion theory at
hand that allows polarized tritium 
$\beta $ decay  to drive the neutrino mass 
closer and closer to zero \cite{Bonn} without  
contradicting observation of possibly larger
neutrino mass in oscillation-- , and
in  $0\nu\beta \beta$ phenomena, thus providing an intriguing and
experimentally testable signature for a potentially
viable and non-trivial  impact
of Majorana spinors on phenomenology.

\section{Acknowledgments.}
One of us (M.K.) benefited from collaboration with 
D.\ V.\ Ahluwalia at the very preliminary stage of this
article and specifically from his knowledge
on the impact relative phases between Weyl spinors in 
$(1/2,0)\oplus (0,1/2)$ may have on phenomenology, 
an idea that he has been tenaciously advocated in 
various of his works.

Work supported by Consejo Nacional de Ciencia y
Tecnologia (CONASyT) Mexico under grant number C01-39820.

\end{document}